\documentclass[a4,aps,amsmath,floatfix,nofootinbib]{revtex4}
\usepackage{graphicx}
\usepackage{color}
\usepackage{enumerate}
\newcommand{\be}{\begin{equation}}
\newcommand{\ee}{\end{equation}}
\newcommand{\ba}{\begin{array}}
\newcommand{\ea}{\end{array}}
\newcommand{\bea}{\begin{eqnarray}}
\newcommand{\eea}{\end{eqnarray}}
\newcommand{\bdm}{\begin{displaymath}}
\newcommand{\edm}{\end{displaymath}}

\begin{document}

\title{Nonequilibrium random-field Ising model on a diluted triangular 
lattice}

\author{Lobisor Kurbah, Diana Thongjaomayum, and Prabodh Shukla}

\affiliation{%
Physics Department \\ North Eastern Hill University \\ 
Shillong-793 022, India}%

\begin{abstract}

We study critical hysteresis in the random-field Ising model (RFIM) on a 
two-dimensional periodic lattice with a variable coordination number 
$z_{eff}$ in the range $3 \le z_{eff} \le 6$. We find that the model 
supports critical behavior in the range $4 < z_{eff} \le 6$, but the 
critical exponents are independent of $z_{eff}$. The result is discussed 
in the context of the universality of nonequilibrium critical phenomena 
and extant results in the field.

\end{abstract}

\maketitle

\section{Introduction}

Disordered systems have been studied extensively over the last few 
decades. A point of special interest is the effect of quenched disorder 
in the system. The quenched disorder endows the system with a large 
number of metastable states that are surrounded by high energy barriers. 
This alters how the system relaxes and how it responds to an external 
driving force. The response becomes sporadic and jerky even if the 
driving force increases slowly and smoothly. The reason is that the 
system is unable to move from one metastable state to another unless the 
applied force has reached the necessary level for crossing the barrier 
between the two metastable states. The passage from one state to another 
involves an avalanche (a quick succession of restructuring events) in 
the internal structure of the system. A large variety of systems exhibit 
this avalanche dynamics. Examples include martensitic transformations in 
many metals and alloys as they are cooled from a bcc structure to a 
closed packed structure~\cite{Eduard1}, earthquakes caused by the 
movement of tectonic plates ~\cite{babcock}, re-arrangement of domains 
in a ferromagnet, and motion in granular materials ~\cite{bretz}. The 
size of an avalanche may vary over a wide range; it may be microscopic, 
large, or critical in the sense of a diverging size. We are particularly 
interested in this paper in nonequilibrium critical phenomena which are 
caused by a diverging avalanche. This is akin to critical phenomena in 
systems in equilibrium that are caused by a diverging correlation 
length.  There is a good deal of experimental support largely from 
Barkhausen noise experiments with a wide variety of 
materials~\cite{meisel,urbach} that nonequilibrium critical phenomena 
have much in common with their equilibrium counterpart such as power 
laws spanning many decades and universal critical exponents.

Apparently there is a significant difference between universality of 
equilibrium critical phenomena and its counterpart in the 
avalanche-driven behavior. In equilibrium, the dimensionality $d$ of the 
system is a key player that determines the universality class of 
critical behavior. Short-range structure of the system is irrelevant for 
this purpose. The rationale is that the critical behavior is caused by 
spontaneous fluctuations in the system whose size diverges at the 
critical point. It should not be influenced by short-range details of 
the system. Thus the critical phenomena on a sc, bcc, or fcc structure 
should be the same. This idea is well tested theoretically as well as 
experimentally in the equilibrium case and it may be expected to hold in 
the nonequilibrium case as well. However we shall see in the following 
that it is not the case. Extensive study of nonequilibrium critical 
behavior has been carried out in the framework of hysteresis in the 
random-field Ising model (RFIM) on a sc lattice with on-site quenched 
random-field having a Gaussian distribution with mean value zero and 
standard deviation $\sigma$ ~\cite{sethna,perkovic1}. The basic quantity 
studied is $m(h,\sigma)$, the magnetization per site in the system at 
zero temperature, as the applied field $h$ is ramped up slowly from 
$h=-\infty$ to $h=\infty$. Extensive numerical studies on the sc 
lattice~\cite{dahmen1,dahmen2,perkovic1,perkovic2} reveal a critical 
value of $\sigma=\sigma_c\approx 2.16 $ such that $m(h,\sigma)$ is 
macroscopically continuous if $\sigma > \sigma_c$ but has a jump 
discontinuity if $\sigma < \sigma_c$. The size of the jump discontinuity 
reduces as $\sigma \rightarrow \sigma_c$ and the jump occurs at larger 
values of the applied field $h$. The point $(\sigma_c=2.16,h_c=1.435)$ 
is a nonequilibrium critical point marked by scale invariant phenomena 
similar to the equilibrium critical phenomena. The model has been 
studied on a square lattice as well ~\cite{frontera,dahmen3,spasojevic}.  
Initial results on the square lattice were inconclusive but the most 
recent studies ~\cite{spasojevic} show the existence of a critical 
point. These results suggest that 2d may be the lower critical 
dimension for the nonequilibrium critical behavior.

The problem of hysteresis in the nonequilibrium random-field Ising model 
at zero temperature can be solved analytically on a Bethe lattice of an 
arbitrary coordination number $z$ ~\cite{dhar}. The exact solution 
brings out a surprising fact that the nonequilibrium critical point 
$(\sigma_c,h_c)$ does not exist if $z<4$. Numerical studies of the model 
on periodic lattices suggest a similar result i.e. the absence of 
avalanche-driven criticality if $z<4$, irrespective of the spatial 
dimensionality $d$ of the lattice ~\cite{sabhapandit1}. For example, 
consider two different $2d$ periodic lattices, (i) the honeycomb lattice 
with $z=3$, and (ii) the triangular lattice with $z=6$. A nonequilibrium 
critical point is absent on the honeycomb lattice ~\cite{sabhapandit1} 
but present on the triangular lattice ~\cite{diana}. The pattern of 
extant results suggests that a lower critical coordination number 
$(z_c=4)$ has greater significance than the idea of a lower critical 
dimension for nonequilibrium critical phenomena. In order to examine 
this point further, we study the problem on a randomly diluted 
triangular lattice. The triangular lattice comprises three equivalent 
inter-penetrating sublattices A, B, and C. We randomly decimate the 
sites on one of these sublattices, say the sublattice C. If $c$ is the 
concentration, i.e. the fraction of sites present on the sublattice C, 
then the limit $c=0$ corresponds to the two sub-lattices A and B making 
up a honeycomb lattice with coordination number equal to 3. For other 
values of the dilution parameter $c$, sites on the lattices A or B have 
an average coordination number $z_{eff}(c)=3(1+c)$. We look for the 
presence of a critical point in the hysterestic response of the system 
for various values of $c$ and find that the critical point disappears if 
$c\le0.33$, or equivalently if the effective coordination number 
$z_{eff}(c)$ is less than 4. We also examine the dependence on $c$ of 
the critical exponent $\nu$ that characterizes the divergence of the 
largest avalanche on the lattice. We conclude that $\nu$ is independent 
of $c$ in the range ($0.33 < c < 1$) within numerical errors.

The paper is organized as follows. In section-II we describe the RFIM 
and its dynamics very briefly for the sake of completeness and to set up 
our notation. Section III presents numerical results for the 
magnetization curve on the lower branch of the hysteresis loop for a 
selected value of dilution ($c=0.90$), and selected values of $\sigma$ 
characterizing the random-field distribution ($\sigma=0.95$, $1.01$, and 
$1.05$). This section emphasizes the key idea for determination of the 
critical point. Section IV contains the body of our numerical results 
for a systematically diluted lattice and finite size analysis. Section V 
contains a brief discussion of the main results of our study.

\section{The Model and Qualitative Behavior}

The RFIM with nearest neighbor ferromagnetic interaction J is 
characterized by the Hamiltonian,
 
\be H=-J 
\sum_{i,j}{c_{i}c_{j}s_{i}s_{j}}-\sum_{i}h_{i}c_{i}s_{i}-h\sum_{i} 
c_{i}s_{i}\ee

Here $c_i$ is a quenched binary variable; $c_i=1$ if site $i$ of a 
$L\times L$ triangular lattice shown in figure (1) is occupied by an 
Ising spin $s_i=\pm{1}$, and $c_i=0$ otherwise. Each site has a quenched 
random-field $h_i$ with a Gaussian distribution of mean value zero and 
standard deviation $\sigma$. The system is placed in a uniform external 
field $h$ which is varied slowly from $h=-\infty$ to $h=\infty$.

The discrete time, single spin flip serial dynamics of the 
zero-temperature nonequilibrium RFIM is specified by the equation 
~\cite{glauber},

\be s_i(t+1)=sign\mbox{ } l_i(t); l_i=J \sum_{j}{c_{j}s_{j}}+h_{i}+h \ee

Figure (1) shows three inter-penetrating sub-lattices A, B and C that 
comprise a triangular lattice. We randomly dilute the sub-lattice C. 
Thus only a fraction c of the sites on the sublattice C are occupied by 
Ising spins. It is easily seen that in the limit $c=0$, the triangular 
lattice reduces to the honeycomb lattice and the coordination number of 
the lattice reduces from 6 to 3. For other values of $c$ in the range $0 
< c < 1$, the lattice is characterized by an inhomogeneous coordination 
number. An occupied site on the sublattice C always has 6 nearest 
neighbors; 3 on sublattice A and 3 on sublattice B. A site on 
sublattice A (B) has 3+3$c$ nearest neighbors; 3 on sublattice B (A) and 
3c (on average) on sublattice C. Thus the coordination number for an 
occupied site on C is 6, and the effective coordination number on A or B 
is given by $z_{eff}($c$)=3(1+$c$)$. The average coordination number 
when A, B, and C are all taken together is $Z_{av} = 6(1+2$c$)/(2+$c$)$. 
In the following we study the magnetization and the character of the 
avalanches on the diluted triangular lattice for different values of 
dilution $c$. The magnetization per site $m(h,\sigma,c;t)$ on the 
lattice is given by

\be m(h,\sigma,c;t)=\frac{1}{L \times L}\sum_i^{L \times L}c_i s_i \ee

We keep $h$ fixed and iterate the single spin flip dynamics till it 
reaches a fixed point i.e. a $t$-independent magnetization. We start 
with $m(h=-\infty,\sigma,c)=-(2+c)/3$ and raise $h$ till some spin 
becomes unstable and flips up. It may cause some of its neighbors to 
flip up as well. The dynamics is iterated till a stable configuration is 
reached. The magnetization of the stable configuration is calculated and 
the process is repeated by raising $h$ to the next higher value that 
makes a new site unstable. The number of spins that flip up in going 
from one fixed point to the next is the size of the avalanche.

The size of an avalanche depends on three quantities; the value of the 
dilution parameter $c$, the standard deviation $\sigma$ of the quenched 
random field, and the applied field $h$. For $c=1$ ~\cite{diana} there 
is a critical value of $\sigma=\sigma_c(1)$ that separates two different 
behaviors that are easy to understand qualitatively. For $\sigma >> 
\sigma_c(1)$, the distribution of random-field is very wide. Therefore 
the spins tend to flip up independently of each other and avalanches are 
relatively small at any applied field $h$. On the other hand, for 
$\sigma << \sigma_c(1)$ the distribution of on-site random fields is 
very narrow. In this case, a single spin flip that changes the local 
field at each nearest neighbor by an amount $2|J|$ causes a spanning 
avalanche across the system if the applied field is greater than some 
threshold. This results in a large change in the magnetization of the 
system, i.e. a first-order jump in the magnetization at some critical 
value of the applied field. The size of the jump decreases with 
increasing $\sigma$ and reduces to zero at the critical point 
$\sigma=\sigma_c(1)$ and $h=h_c(1)$. We call it a nonequilibrium 
critical point because the hysteretic susceptibility of the system 
diverges at this point. We find that a similar critical point is present 
on the diluted lattice in the restricted range $0.33 < c <1$. The value 
of $\sigma_c(c)$ reduces with decreasing $c$ and drops discontinuously 
to zero at $c=0.33$ approximately. As mentioned earlier $c=1/3$ 
corresponds to $z_{eff}=4$. Thus a critical point occurs on the diluted 
lattice only if $z_{eff} >4$. This is interesting in the context of a 
similar conclusion reached in the study of the Bethe lattice as well as 
some other periodic lattices.

The central object of the present study is the determination of 
$\sigma_c(c)$. It seems nearly impossible to determine it analytically. 
This is not surprising given that exact solutions of Ising models, 
particularly with quenched disorder, are rare.  We have to resort to 
numerical simulations of the model. Simulations too have difficulties of 
choosing a good algorithm that suits available computing resources. For 
each $c$, we first determine a range $\sigma_{min}<\sigma<\sigma_{max}$ 
that may contain $\sigma_c(c)$. This is done by locating $\sigma_{max}$ 
where no avalanche on the hysteresis loop is macroscopic, and a 
$\sigma_{min}$ where there is clearly a macroscopic jump in the 
magnetization. The next step is to study the system for different values 
of $\sigma$ in the range $[\sigma_{min},\sigma_{max}]$ at suitably 
chosen intervals $\delta\sigma$ with a view to narrow this range and 
locate the critical point that separates the two behaviors. There are 
two difficulties. For each value of $\sigma$, the avalanche distribution 
is determined by taking an average over a large number of 
configurations. It is a cpu intensive exercise and a compromise is 
necessary in choosing an optimal $\delta\sigma$. The size of the step 
$\delta\sigma$ is one factor that contributes to the error bars in the 
results. The other difficulty is that fluctuations increase as we 
approach the vicinity of the critical point. The avalanches that were 
microscopic for $\sigma >> \sigma_c$ grow in size and compete with the 
avalanches associated with a jump in magnetization even as the jump 
approaches zero. We need a method to distinguish the infinite avalanche 
associated with a first order jump in magnetization from an infinite 
avalanche associated with a critical point. These two categories of 
avalanches are distinguished from each other by looking at their 
probability distributions. This is again a cpu intensive process. 
Qualitatively the largest avalanche associated with a magnetization jump 
scales linearly with the system size $L$ while the largest avalanche 
associated with a critical point scales as $L^{1/2}$. The shapes of the 
avalanche distribution functions for $\sigma<\sigma_c$ and $\sigma 
\approx \sigma_c$ are different from each other~\cite{diana,farrow}. 
This difference is exploited to pin down $\sigma_c$ within error bars as 
described in the following.

\section{Simulations}

The shape of magnetization curve $m(h,\sigma,c)$ in a changing field $h$ 
depends on the starting state of the system in addition to $h$, 
$\sigma$, and $c$. We start from $m(h=-\infty,\sigma,c)=-(2+c)/3$, and 
raise the field slowly to $h=\infty$. Figure (2) shows $m(h,\sigma,c)$ 
on a $6000\times 6000$ lattice for $c=0.90$, and three different values 
of $\sigma=0.95$ (blue triangles), $\sigma=1.01$ (black circles), and 
$\sigma=1.05$ (red squares) in the range $1.5 < h < 1.7$. It illustrates 
three categories of behavior: (i) at smaller values of $\sigma$ (blue 
triangles) $m(h,\sigma,c)$ has a jump discontinuity, (ii) at larger 
values of $\sigma$ (red squares) $m(h,\sigma,c)$ looks apparently smooth 
if we allow for fluctuations due to finite size effects that are 
naturally present in any numerical simulation, (iii) there is an 
intermediate region (black circles) where it is relatively difficult to 
decide if the curve is smooth or has a discontinuity. Somewhere in this 
region lies a critical value $\sigma_c$ with the corresponding 
magnetization curve being theoretically smooth but containing a point of 
inflexion at $h=h_c$. The inflexion point is the critical point where 
the fluctuations in the system become anomalously large and consequently 
the susceptibility of the system diverges. The difficulty is that it is 
not easy to identify the inflexion point in simulations. Fluctuations 
also increase with increasing size of the system, and simulated 
trajectories remain qualitatively similar to the ones shown in figure 
(2). Thus it is unreasonable to expect that simulation with a larger 
system (which would anyway require an unreasonably long computer time) 
would make it any easier to locate the inflexion point by merely looking 
at the curve. Rather it has to be inferred indirectly from the analysis 
of fluctuations in its vicinity. The fluctuations are anomalously large 
not only at the critical point $(h_c,\sigma_c)$ but also at the 
discontinuity in $m(h,\sigma,c)$ at $\sigma < \sigma_c$ and $h<h_c$. 
This presents an additional difficulty in determining the critical 
point. Our approach ~\cite{farrow,diana} is based on the idea that the 
character of fluctuations at the critical point is different from the 
character of fluctuations at a first order discontinuity. The 
macroscopic discontinuity in magnetization constitutes a large avalanche 
but avalanches elsewhere on the magnetization trajectory are 
exponentially small. Thus on a logarithmic scale, the probability 
$P(s,\sigma,c)$ of an avalanche of size $s$ has two parts, a part that 
decreases linearly with $s$ and another a delta function peak at $s 
\approx s_{max}$; $s_{max}$ is of the order of the system size but 
decreases as $\sigma$ approaches $\sigma_c$ from below. Also the 
distribution of avalanches is asymmetric on the two sides of the 
discontinuity. Avalanches tend to be larger at the onset of the 
discontinuity. After the discontinuity has occurred, most of the spins 
in the system have turned up and the number of potential sites that can 
turn up reduces drastically. Thus avalanches immediately after a jump in 
magnetization tend to be smaller in comparison to those just before it. 
In contrast to this, avalanches on both sides of a critical point are 
similar to each other. The maximum size of a critical avalanche scales 
as the square root of the system size while it scales linearly with the 
system size at a discontinuity. However in spite of these distinguishing 
features between a first-order and a second order transition, the issue 
still remains difficult to decide because as $\sigma \rightarrow 
\sigma_c$ from below, the size of the first order jump tends to zero and 
the Delta function peak at $s=s_{max}$ tends to vanish. In the absence 
of a more efficient method, we use the same (rather laborious) method to 
locate critical points on the diluted triangular lattice as was used on 
the undiluted lattice ($c=1$). In the following we discuss the case 
$c=0.90$ in detail, and present the results for other values of dilution 
in the form of a table and graphs.

As discussed in reference~\cite{diana}, we count the occurrence of all 
avalanches of size $s$ as the system is driven from $h=-\infty$ to 
$\infty$. Let $P(s,\sigma,c)$ be the probability of an avalanche of size 
$s$ anywhere on the $m(h,\sigma,c)$ curve in an increasing applied field 
($h=-\infty$ to $h=\infty$) on a triangular lattice whose sites on 
sublattice C are occupied with probability $c$. As mentioned earlier, 
$\log{P(s,\sigma,c)}$ will have a linearly decreasing part in the range 
$1 < s < s_{max}$ if $m(h,\sigma,c)$ has a discontinuity. This is born 
out by the red (filled circles) curve in figure (3) which shows the raw 
data for $\sigma=1.1$ on a $240 \times 240$ lattice for $c=0.90$ and 
$50000$ independent realizations of the random-field distribution. The 
green (filled squares) curve in the same figure shows the data 
for $\sigma=1.295$ which we estimate to be the critical value. 
Simulations were performed for several closely spaced values of 
$\sigma$. However we show the data for only three values of $\sigma$ so 
as not to crowd figure (3). The data depicted in green is the closest to 
a linear decrease. The Delta function peak has vanished signifying that 
the jump in the magnetization has approached zero. Data for $\sigma=1.5$ 
is shown in blue (filled triangles). We conclude that the value of 
$\sigma$ corresponding to the blue must be greater than $\sigma_c$ due 
to two reasons: (i) the largest avalanche $s_{max}$ with a non-zero 
probability of occurrence is far less than the system's size, and (ii) 
$P(s,\sigma,c)$ does not approach to zero linearly at $s_{max}$ but 
rather bends down to it.

Figure (4) shows similar data as figure (3) but in a binned form. 
Binning reduces the scatter of the data and we are able to show the data 
for a larger set of closely spaced $\sigma$ values without crowding the 
figure too much. The raw data for each value of $\sigma$ has been binned 
in 50 linear bins. Here the values of $\sigma$ are very close to 
$\sigma_c$ so that the largest avalanche in each case is of the same 
order of magnitude. What distinguishes different values of $\sigma$ is 
that curves that bend up near the largest avalanche indicate a tendency 
to form a Delta function peak. These indicate that the corresponding 
$\sigma$ is smaller than $\sigma_c$. Similarly the curves that bend down 
near the largest avalanche indicate that the corresponding values of 
$\sigma$ are larger than $\sigma_c$. The curve closest to a straight 
line (blue circles) belongs to $\sigma_c=1.295$. Before closing this 
section, two comments on the binning procedure may be in order. We have 
also used logarithmic binning which is normally the preferred binning 
procedure when the distributions have a fat tail as in our case. This 
does not change the results presented here. The main purpose of figure 
(4) is to distinguish between curves that bend up near the end from 
those that bend down. This is easier for the eye if linear binning is 
used due to higher density of data points in the region. The second 
point is that the last bin in each case should be ignored due to lack of 
sufficient number of data points in this bin.

\section{Finite Size Effects and Results}

Using the procedure outlined above, we have determined the critical 
value $\sigma_c(L,c)$ on diluted $L \times L$ lattices for various 
values of $L$ in the range 100 to 600. These are rather small sizes 
compared with the ones used in the study of the undiluted problem on a 
square lattice~\cite{dahmen3,spasojevic}. However, such small sizes were 
found to be adequate in reference~\cite{diana} to demonstrate the 
presence of critical hysteresis on a triangular lattice. In the present 
study as well, these appear to be adequate to examine the effect of 
dilution parameter $c$ and the effective coordination number $z_{eff}$ 
on critical hysteresis in the system.  Our results are presented in 
Table I. According to the scaling hypothesis, the correlation length in 
the vicinity of the critical point scales as $\sim [\sigma_c(c) 
-\sigma]^{-\nu(c)}$ as $\sigma \rightarrow \sigma_c(c)$ from below. The 
argument on $\nu(c)$ indicates that we are open to the possibility that 
the the critical exponent $\nu$ may depend on the amount of dilution on 
the lattice. However, as we shall see in the following, this turns out 
not to be true within the error bars of our analysis. In the present 
problem, the correlation length is measured by the largest distance that 
an avalanche travels from its point of origin. On a finite lattice, the 
farthest distance an avalanche can travel is limited by the size of the 
lattice. Thus $L \sim [\sigma_c(L,c) -\sigma]^{-\nu(c)}$ where 
$\sigma_c(L,c)$ is a lattice-dependent critical value of $\sigma_c(c)$.  
Allowing for a constant of proportionality, we may 
write~\cite{spasojevic}

\be L^{-\frac{1}{\nu(c)}}=\frac{\sigma_c(L,c)-\sigma_c(c)}{\sigma_c(c)} 
\mbox{ or } -\frac{1}{\nu(c)} \log_{10} L = \log_{10} \left[ 
\frac{\sigma_c(L,c)}{\sigma_c(c)} -1\right]\ee

Here the factor $1/\sigma_c(c)$ appearing immediately after the first 
equality sign is a constant of proportionality. A more general constant 
of proportionality $1/a$ is considered in equation (5).

\begin{table}[h] \caption{$\sigma_c(L,c)$} \centering 
\begin{tabular}{c| c c c c c c c c c} \hline\hline L & ${c=1.00}$ & 
${c=0.90}$ & ${c=0.80}$ & ${c=0.70}$ & ${c=0.60}$ & ${c=0.50}$ & 
${c=0.40}$ & ${c=0.34}$ \\ [0.5ex] \hline 99 & 1.63 $\pm 0.01$ & 
1.47$\pm 0.01$ & 1.305$\pm 0.005$ & 1.125 $\pm 0.005$ & 0.885$\pm 0.005$ 
& 0.665$\pm 0.005$ & 0.51$\pm 0.01$ & 0.485$\pm 0.005$\\ [3ex]

120  & 1.59 $\pm 0.01$  & 1.425$\pm 0.005$ & 1.255$\pm 0.005$  & 1.065 $\pm 0.005$
 & 0.83$\pm 0.01$ & 0.62$\pm 0.01$ & 0.485$\pm 0.005$ & 0.46$\pm 0.01$\\ [3ex]

141  & 1.56 $\pm 0.01$  & 1.39$\pm 0.01$ & 1.215$\pm 0.005$  & 1.02 $\pm 0.01$
 & 0.785$\pm 0.005$ & 0.585$\pm 0.005$ & 0.465$\pm 0.005$ & 0.44$\pm 0.01$\\ [3ex]

168  & 1.525 $\pm 0.005$  & 1.355$\pm 0.005$ & 1.175$\pm 0.005$  & 0.98 $\pm 0.01$
 & 0.745$\pm 0.005$ & 0.555$\pm 0.005$ & 0.445$\pm 0.005$ & 0.425$\pm 0.005$\\ [3ex]

198 & 1.50$\pm 0.01$   & 1.325$\pm 0.005$  & 1.14$\pm 0.01$  & 0.94$\pm 0.01$
 & 0.71$\pm 0.01$ & 0.525$\pm 0.005$ & 0.425$\pm 0.005$ & 0.405$\pm 0.005$\\ [3ex]
 
240 & 1.47$\pm 0.01$   & 1.295 $\pm 0.005$  & 1.105$\pm 0.005$  & 0.90$\pm 0.01$
 & 0.675$\pm 0.005$ & 0.495$\pm 0.005$ & 0.41$\pm 0.01$ & 0.39$\pm 0.01$\\ [3ex]

300 & 1.44$\pm 0.01$ & 1.26$\pm 0.01$  & 1.065$\pm 0.005$  & 0.86$\pm 0.01$
 & 0.635$\pm 0.005$ & 0.465$\pm 0.005$ & 0.39$\pm 0.01$ & 0.375$\pm 0.005$\\ [3ex]

360 & 1.42$\pm 0.01$ & 1.235$\pm 0.005$  & 1.04$\pm 0.01$  & 0.83$\pm 0.01$
 & 0.605$\pm 0.005$ & 0.445$\pm 0.005$ & 0.375$\pm 0.005$ & 0.365$\pm 0.005$\\ [3ex]

390 & 1.41$\pm 0.01$ &  1.225$\pm 0.005$  & 1.03$\pm 0.01$  & 0.815$\pm 0.005$  
& 0.595$\pm 0.005$ & 0.435$\pm 0.005$ & 0.37$\pm 0.01$ & 0.36$\pm 0.01$\\ [3ex]

480 & 1.39$\pm 0.01$ &  1.20$\pm 0.01$  & 1.00$\pm 0.01$  & 0.785$\pm 0.005$ 
& 0.565$\pm 0.005$ & 0.415$\pm 0.005$ & 0.355$\pm 0.005$ & 0.345$\pm 0.005$\\ [3ex]

600 & 1.37$\pm 0.01$  & 1.18$\pm 0.01$  & 0.975$\pm 0.005$   & 0.76$\pm 0.01$
 & 0.54$\pm 0.01$ & 0.395 $\pm 0.005$ & 0.345$\pm 0.005$ & 0.335$\pm 0.005$\\ [3ex]
\hline\hline
\end{tabular}
\end{table}

For each value of dilution $c$, the data in Table I is used to plot 
$-\log_{10} L$ vs. $\log_{10} [\sigma_c(L,c)/\sigma_c(c)-1]$ for 
different values of parameter $\sigma_c(c)$. The shape of the plot 
changes from concave up to a straight line, and then to concave down as 
$\sigma_c(c)$ is increased from zero. We search for the best value of 
the $\sigma_c(c)$ that produces a straight line. However, the transition 
from concave up to concave down is not sharp and spreads over a broad 
range. This creates a relatively large uncertainty in the value of 
$\sigma_c(c)$ that produces the best fit, and also an uncertainty in the 
slope of the line. For example, the best fit to a straight line for the 
data in the second column ($c=0.90$) is obtained for $\sigma_c=1.00 \pm 
0.04$. This is shown in figure (5). The slope of the straight line is 
equal to $1/\nu(c)=0.54 \pm 0.07$, or $\nu(c)=1.85 \pm 0.26$. Similarly, 
we determine $\nu(c)$ for other values of dilution $c$. Our results for 
$\sigma_c(c)$ and $\nu(c)$ for different values of dilution on the 
triangular lattice are summarized in the following table.

\begin{table}[h]
\caption{$\sigma_c(c)$ and $\nu(c)$ for different values of dilution $c$}
\centering
\begin{tabular}{c| c c c c c c c c c}
\hline\hline
& ${c=1.00}$ &  ${c=0.90}$ & ${c=0.80}$ & ${c=0.70}$ & 
${c=0.60}$ & ${c=0.50}$ & ${c=0.40}$ & ${c=0.34}$ \\ [0.5ex]
\hline
$\sigma_c $ & \hspace{.5cm} 1.22${\pm 0.04}$  & 1.00${\pm 0.04}$ & 0.77${\pm 0.05}$ 
& 0.54${\pm 0.05}$  & 0.33${\pm 0.05}$  & 0.26${\pm 0.04}$ & 0.25${\pm 0.03}$ & 0.25${\pm 0.03}$ \\[3ex]

$\nu(c) $ & \hspace{.5cm} 1.78${\pm 0.29}$  & 1.85${\pm 0.26}$ & 1.87${\pm 0.29}$ 
& 1.83${\pm 0.26}$  & 1.85${\pm 0.28}$  & 1.64${\pm 0.28}$ & 1.76${\pm 0.35}$ & 1.79${\pm 0.33}$ \\[3ex]

\hline\hline
\end{tabular}
\end{table}

Our simulations reveal a critical value $c_c$ of the dilution parameter; 
$c_c=0.33$ approximately. At $c=c_c$, $\sigma_c(c)$ drops to zero 
abruptly, and remains zero for $c < c_c$ as shown in Figure (6). There 
is no first order jump in $m(h,\sigma,c)$ for $c<c_c$. We have verified 
this directly from the simulations as well as inferred it from the 
following. For $c< 0.33$, the plot $-\log_{10} L$ vs. $\log_{10} 
[\sigma_c(L,c)/\sigma_c(c)-1]$ is never concave up for any 
$\sigma_c(c)>0$. This indicates the absence of a disorder driven jump in 
the magnetization and therefore the disappearance of a critical point if 
$c<0.33$. To further validate this point, we also tried as in 
ref~\cite{diana}, the scaling form

\be L^{-\frac{1}{\nu(c)}} = \frac{\sigma_c(L,c)-\sigma_{c}(c)}{a}, \ee

\noindent where $a$ is an arbitrary parameter. The role of $a$ is to shift the 
curve along the $y$-axis without changing its shape. If we set 
$\sigma_c(c)=0$ in the above equation, we find that the plot is always 
concave up for any value of $c>c_c$. This means that for $c>c_c$, the 
system possesses a critical point with $\sigma_c>0$. On the other hand, 
for $c<c_c$, the curve is a straight line indicating the absence of a 
critical point.

The variation of $\sigma_c(c)$ vs. $c$ has been plotted in figures (6), 
and $\nu(c)$ vs $c$ in figure (7). We may draw the following conclusions 
from these figures. As the lattice is diluted increasingly, it continues 
to support a critical point but the value of $\sigma_c(c)$ decreases. 
This is intuitively reasonable because the random dilution of the 
lattice amounts to a positional disorder in the system that supplements 
to the disorder due to the random-field. Thus the critical point on the 
diluted lattice corresponds to a narrower distribution of the 
random-field as compared with the undiluted case. We also note that the 
critical dilution $c_c=0.33$ corresponds to $z_{eff} \approx 4$.

Figure (7) shows that the exponent $\nu(c)$ is independent of $c$ within 
the error bars. We may expect this universality to hold for other 
exponents as well because of the relationship between different 
exponents. It is unlikely that only one exponent in an equation is 
universal while others are not. We have tried to extract a bit more 
information from our numerical data regarding other exponents. We have 
tried the collapse of the integrated avalanche size distribution with 
the following scaling form ~\cite{perkovic2,dahmen3}:

\be P(S,\sigma) \sim S^{-(\tau +\alpha \beta \delta)} 
\tilde{P}(S|r|^{1/\alpha}) \ee

Here $\alpha$ is the exponent describing the largest cut-off avalanche 
i.e. $S_{max} \sim|r|^{-1/\alpha}$ where 
$r=\frac{\sigma-\sigma_{c}}{\sigma}$, $\tau$ is the avalanche size 
exponent, $\beta$ gives the scaling of the change in magnetisation due 
to spanning avalanche, $\Delta m \sim |r|^\beta$, at the critical field. 
The exponent $\delta$ describes the scaling of the reduced magnetisation 
with reduced magnetic field, $m\sim h^{\delta}$ at $\sigma_c$. Thus the 
product $P(S,\sigma) \times S^{(\tau +\alpha \beta \delta)}$ is a 
function of a single variable $S|r|^{1/\alpha}$. Fig(8) shows 
$\tilde{P}(S|r|^{1/\alpha})$ vs. $S|r|^{1/\alpha}$ for $c=0.80$; $L=168, 
240$ and $390$; and different values of $\sigma$ ranging from $1.15$ to 
$1.40$. The curves collapse on each other reasonably well if we choose 
$\tau+\alpha\beta\delta=2.05$ and $\alpha=0.12$.

We have also examined how the exponents extracted from the collapsing 
curves depend on $L$ and $|r|$. For this purpose, we fix $c$ and $L$ 
(say $c=0.80$, and $L=390$), and choose three closely spaced values of 
$\sigma$ (say $\sigma=1.05, 1.06,$ and $1.07$). The three values of 
$\sigma$ correspond to three closely spaced values of $|r|$. Let 
$|r|_{avg}$ denotes the average of the three $|r|$ values. We search for 
the values of $\tau+\alpha\beta\delta$ and $\alpha$ that produce the 
best collapse of the three curves associated with $|r|_{avg}$. Next we 
choose several different triplets of closely spaced $\sigma$ values and 
determine the exponents over a range of values of $|r|_{avg}$. This 
exercise is repeated for systems of different sizes ($L=168, 240, 390, 
600$), and for a range of values of $c$ in the range $c > 0.33$. The 
results for $c=0.80$ are shown in figures (9) and (10) along with the 
range of variations (error bars) in the values of the exponents. These 
figures also show the extrapolated values of the exponents in the limit 
$|r|_{avg} \rightarrow 0$ or $L \rightarrow \infty$. Thus we obtain 
$\tau +\alpha\beta\delta = 2.05\pm 0.05$ and $\alpha=0.11 \pm 0.02$ for 
$c=0.80$. We have checked that the values of the exponents for other 
values of dilution $c$ lie in the same range as for $c=0.80$. This leads 
us to conclude that the values of the exponents $\tau 
+\alpha\beta\delta$ and $\alpha$ are independent of $c$ if $c > 0.33$.

\section{Discussion}

We have investigated critical hysteresis on a two dimensional lattice 
with a variable coordination number. We had to study the problem 
numerically because exact analytical solutions of this problem are not 
possible. Simulations of the model suggest that the three parameters of 
the model $c$, $h$, and $\sigma$ are characterized by critical values 
$c_c$, $h_c$, and $\sigma_c$ respectively. If $c>c_c$, the critical 
point $(h_c,\sigma_c)$ is marked by a diverging avalanche and a critical 
exponent $\nu(c)$. Within numerical errors, we find that the exponent $\nu$ 
does not depend on $c$. The critical value $c_c$ is approximately equal 
to $c_c=1/3$ which corresponds to $z_{eff}=4$ and $Z_{av}=30/7$. We note 
that each site on A or B sublattice has three nearest neighbor sites on 
the C sublattice which are occupied independently with probability $c$. 
Therefore if $c<1/3$, the probability of a spanning path on A+B through 
occupied sites on C goes to zero. In this case the C sublattice does not 
contribute to a diverging correlation length. The cooperative behavior 
of the system is qualitatively the same as on the honeycomb lattice 
comprising A and B, and the relevant parameter is $z_{eff}$ rather than 
$Z_{av}$.

We note that $c=1/3$ corresponds to an effective coordination number 
$z_{eff}=4$. Thus our results suggest that critical behavior disappears 
when the coordination number of the lattice drops below four. A similar 
result holds for the Bethe lattice~\cite{dhar} and also some periodic 
lattices~\cite{sabhapandit1}. On the Bethe lattice, the problem can be 
solved analytically and therefore the mathematical reason for the 
absence of critical behavior on Bethe lattice of coordination number 
less than four is understood. However the physical reason for this 
behavior on periodic lattices is not well understood. It is curious that 
the lower critical coordination number is also equal to four on the 
diluted triangular lattice. We could not have anticipated this result 
beforehand. Our numerical results only suggest $z_{eff}$ to be 
approximately four within errors but the probabilistic argument 
mentioned above indicates that it may be exactly equal to four. Thus all 
the extant results indicate that nonequilibrium critical behavior occurs 
only on lattices with coordination number equal to four or more 
irrespective of the dimensionality of the lattice.

Before closing, we wish to comment on the small difference between 
results presented here for the case $c=1$, and those in reference (16). 
We find $\sigma_c=1.22 \pm 0.04$ and $\nu=1.78 \pm 0.29$ (for c=1.00) 
while the values reported in reference (16) are $\sigma_c=1.27$ and 
$\nu=1.6 \pm 0.2$. The results are consistent with each other within 
error bars. As authors of reference (16) are also coauthors of the 
present paper, we can point out the reason for the difference in the two 
sets of results. We estimate $\sigma_c$ as the best value of this 
parameter that results in a straight line when the left-hand-side of 
equation (4) is plotted against its right-hand-side. As explained in the 
previous section, the range of values of $\sigma_c(c)$ that produces an 
apparent straight line is rather broad. The results in reference (16) 
rely on a visual scrutiny of the plots and choosing the fit that looks 
best to the eye. This is of course susceptible to human error. In the 
present study, we have used the linear least squares fitting technique. 
It has the convenience of a mechanical method to handle the data 
contained in Table I but it also has the disadvantage that outlying 
points have a disproportionate effect on the fit. This contributes to 
the uncertainty in $\sigma_c(c)$ and $\nu$ which is determined by the 
slope of the straight line. The uncertainties in our results are 
relatively large for the effort put in this study. This seems 
unavoidable with the present method. The remark on the closeness of the 
exponent $\nu$ on the triangular lattice ~\cite{diana} and the simple 
cubic lattice ~\cite{perkovic2} should also be taken in the same vein. 
At a qualitative level, it does suggest that the avalanche-driven 
exponent may depend on the coordination number rather than the 
dimensionality of the lattice. However, it is difficult to reach a 
stronger conclusion with the error bars in our analysis. We hope the 
present study will motivate further studies of the issues raised here 
with new and improved techniques.

\begin{figure}[p] 
\includegraphics[width=0.75\textwidth,angle=0]{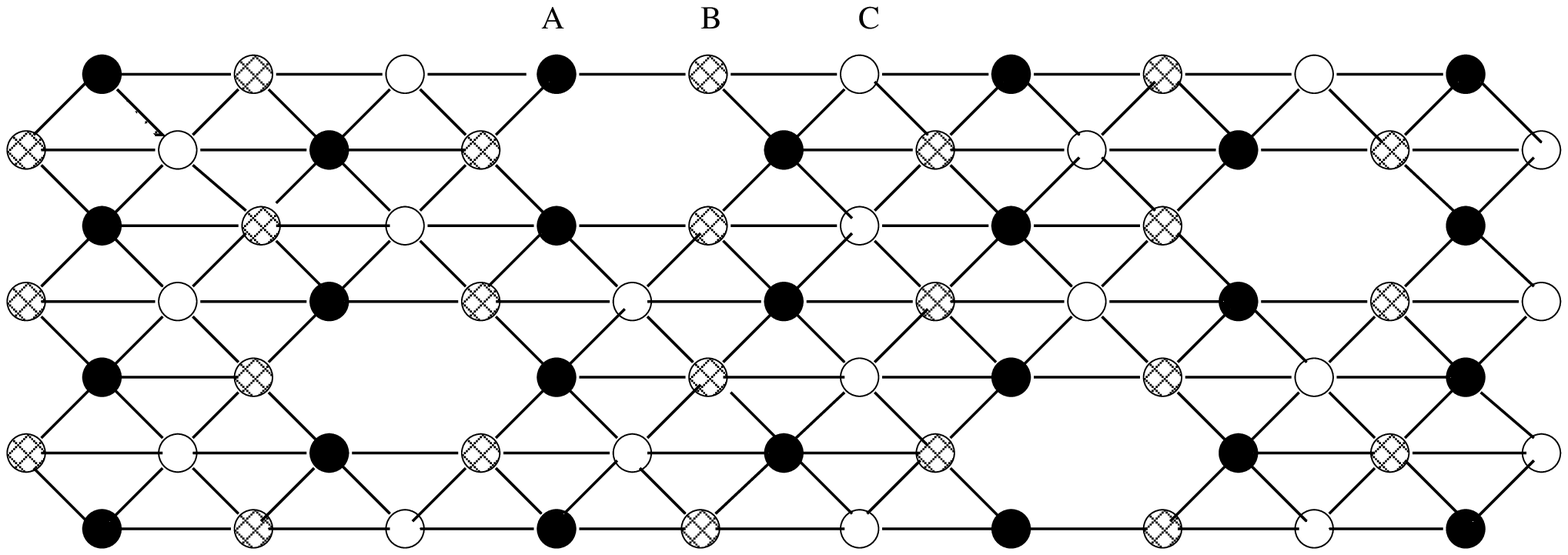} 
\caption{Triangular lattice with sublattices $A $(black circles), $B$ 
(shaded circles) and $C $(white circles). $C $ is randomly diluted.} 
\label{fig1} \end{figure}

\begin{figure}[p]
\includegraphics[width=0.75\textwidth,angle=0]{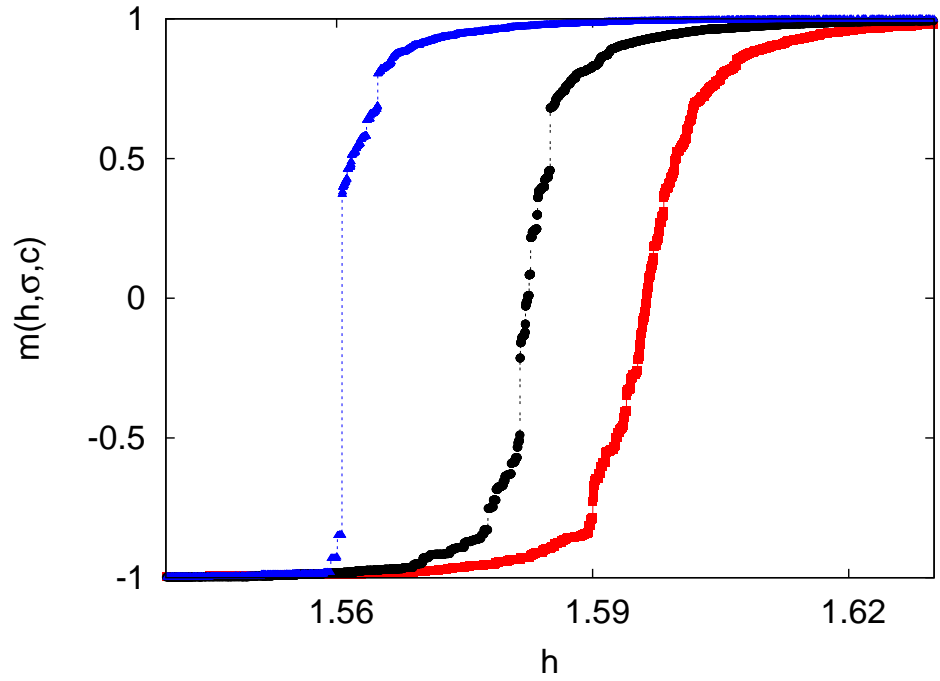}

\caption{(color online) Magnetization curves on a $6000\times 6000$ 
diluted triangular lattice with ${c=0.90}$ and for $\sigma=0.95 $(blue 
triangles), $\sigma=1.01 $(black circles), and $\sigma=1.05$(red squares).}

\label{fig2}
\end{figure}

\begin{figure}[p] 
\includegraphics[width=0.75\textwidth,angle=0]{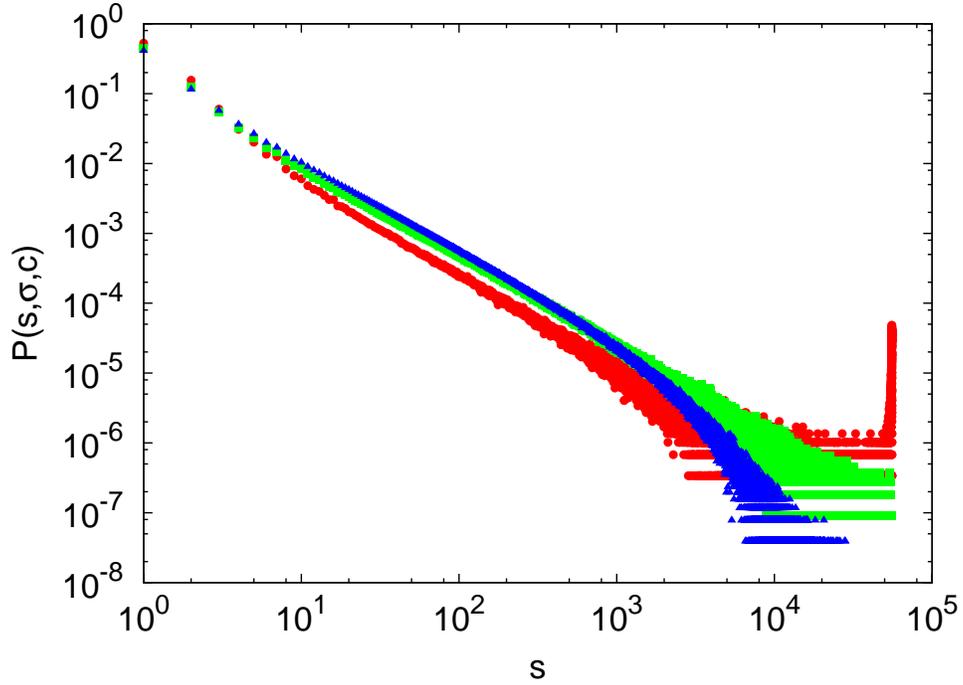} \caption{(color 
online) Distribution of avalanches $P(s,\sigma,c)$ on a $L \times L$ 
lattice for $L=240$ and $c=0.90$; $\sigma=1.10$ (red circles with a peak 
at $s\approx L\times L$), $\sigma=1.295$ (green squares, note the peak at 
$s\approx L\times L$ has vanished and the plot is almost linear) and 
$\sigma=1.50$ (blue triangles, the bending of the curve indicates the largest 
avalanche is smaller than $L\times L$).} \label{fig3} \end{figure}

\begin{figure}[p] 
\includegraphics[width=0.75\textwidth,angle=0]{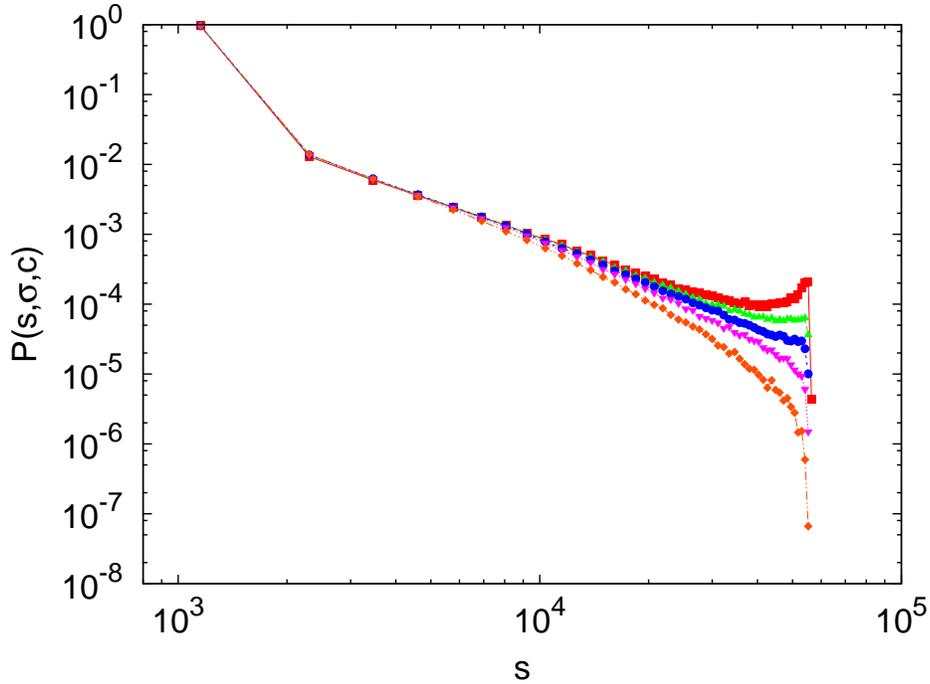} \caption{(color 
online) $P(s,\sigma,c)$ for $c=0.90$ and $L=240$ using linear binning; 
$\sigma=1.25$(red squares), $\sigma=1.27$(green triangles), $\sigma=1.295$(blue circles), 
$\sigma=1.31$(pink inverted triangles), and $\sigma=1.35$(brown diamonds). Figure shows 
$\sigma_{c}(L,c)\approx 1.295$ because the corresponding curve is nearly 
linear; the curves for $\sigma <\sigma_{c}(L,c)$ tend to peak at the 
largest avalanche while those for $\sigma>\sigma_{c}(L,c)$ tend to bend 
down.} \label{fig4} \end{figure}

\begin{figure}[p] 
\includegraphics[width=0.75\textwidth,angle=0]{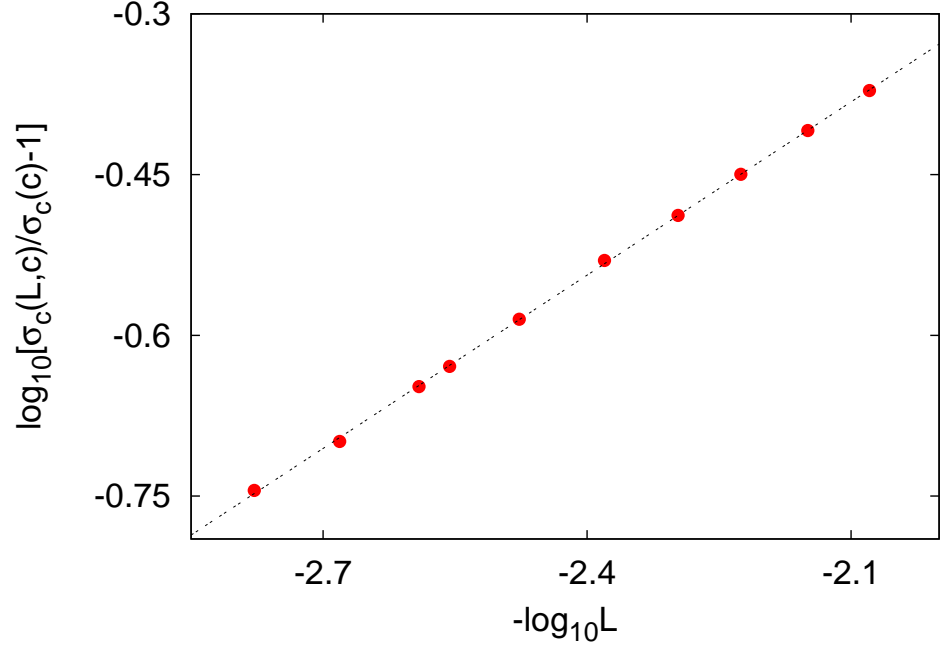} 
\caption{(color online) Plot of 
log$_{10}[\frac{\sigma_{c}(L,c)}{\sigma_{c}(c)}-1] $ vs. -log$_{10}$ $L$ for 
$c=0.90$. The best fit to a straight line is obtained for 
$\sigma_{c}(c)=1.00 \pm 0.04$. The slope of the straight line yields 
$\nu(c)=1.85 \pm 0.26$.} \label{fig5} \end{figure}

\begin{figure}[p] 
\includegraphics[width=0.75\textwidth,angle=0]{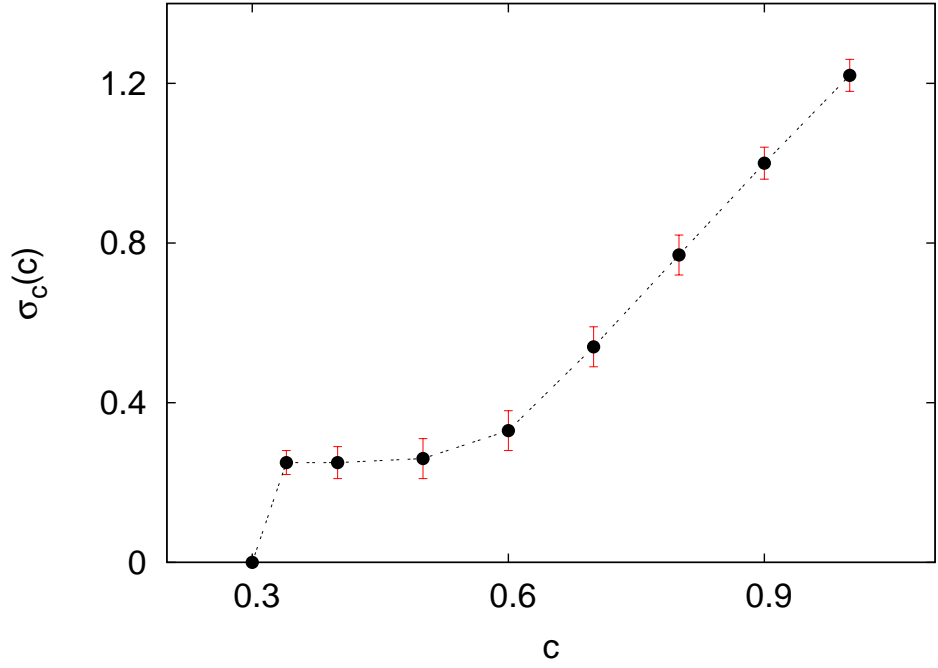} 
\caption{(color online) Variation of $\sigma_{c}(c)$ with $c$. The 
figure shows the absence of critical hysteresis if $c<0.33$.} 
\label{fig6} \end{figure}

\begin{figure}[p]
\includegraphics[width=0.75\textwidth,angle=0]{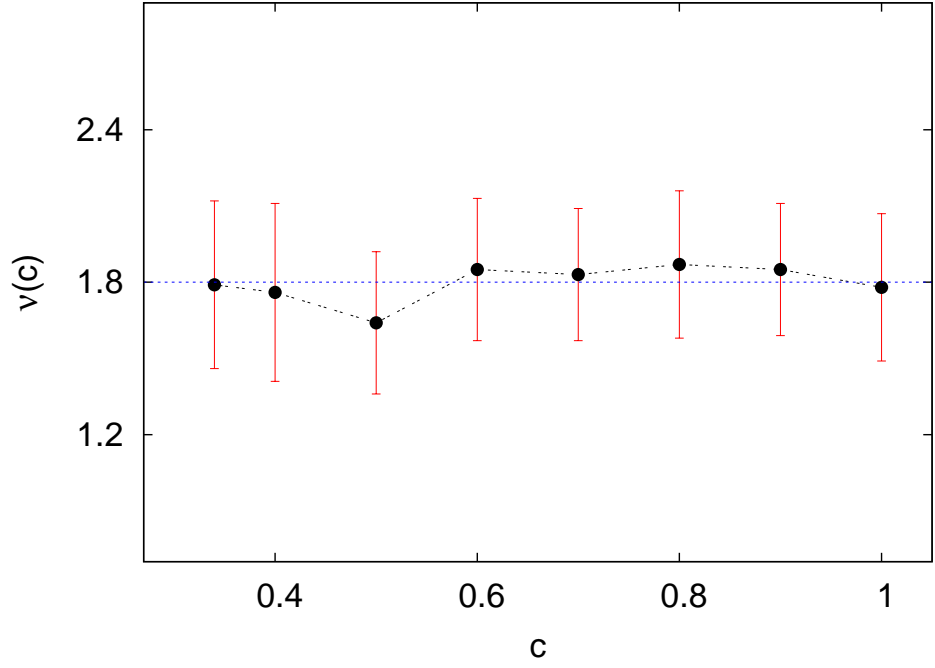}

\caption{(color online) Variation of the exponent $\nu(c)$ with $c$; 
$\nu(c)\approx 1.8\pm 0.3$ for $c>0.33$ } ¬

\label{fig7}
\end{figure}

\begin{figure}[p]
\includegraphics[width=0.75\textwidth,angle=0]{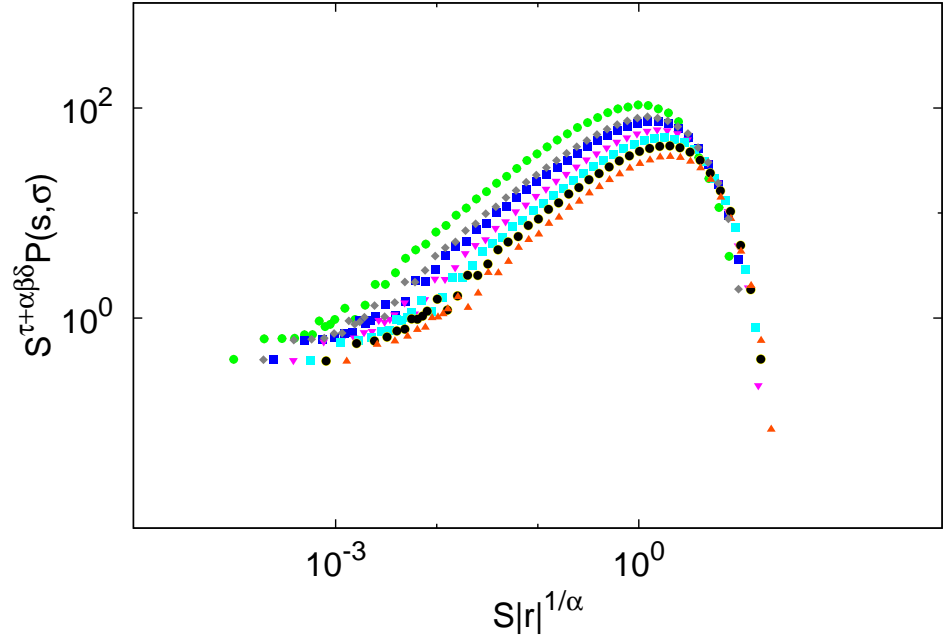}

\caption{(color online) Scaling collapse of the integrated avalanche 
size distribution for $c=0.80$; $L=168, 240,$ and $390$ for seven values 
of $\sigma$ chosen from the range $1.15-1.40$. The best collapse is 
obtained for $\tau+\alpha\beta\delta=2.05$ and $\alpha=0.12$.} ¬

\label{fig8}
\end{figure}

\begin{figure}[p]
\includegraphics[width=0.75\textwidth,angle=0]{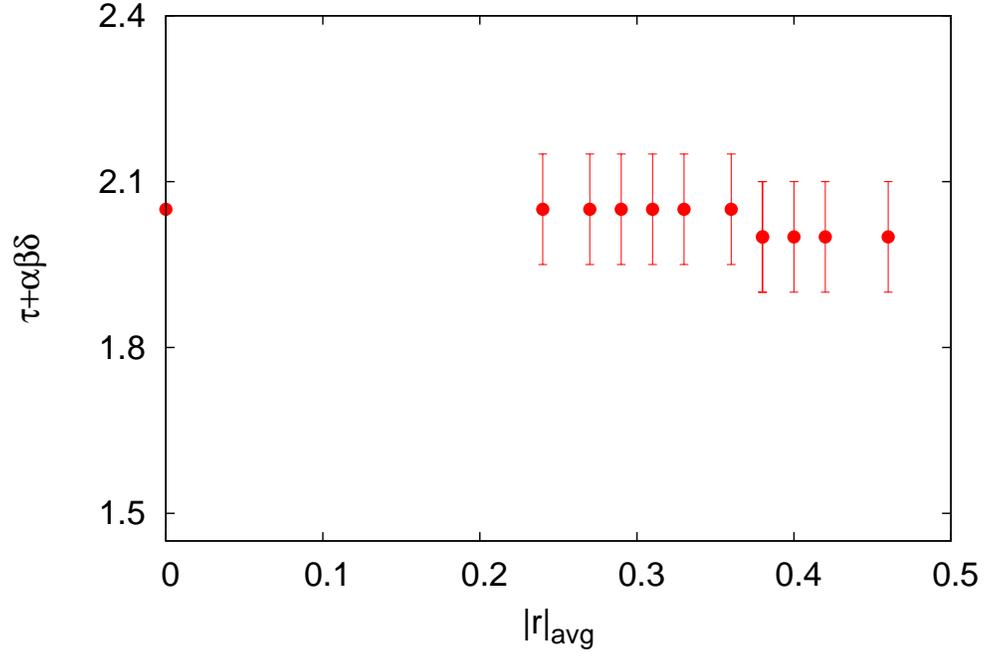}

\caption{(color online) Plot of $\tau+\alpha\beta\delta$ vs. $|r|_{avg}$ 
for $c=0.80$. The values of the exponent are extracted from the best 
collapse of the avalanche size distribution for each set of three 
$\sigma$ values ranging from $\sigma=1.00-1.44$ and different values of 
$L=168,240,390$ and $600$. The extrapolated value is the point at 
$|r|_{avg}\rightarrow 0$.
} 
¬

\label{fig9}
\end{figure}

\begin{figure}[p]
\includegraphics[width=0.75\textwidth,angle=0]{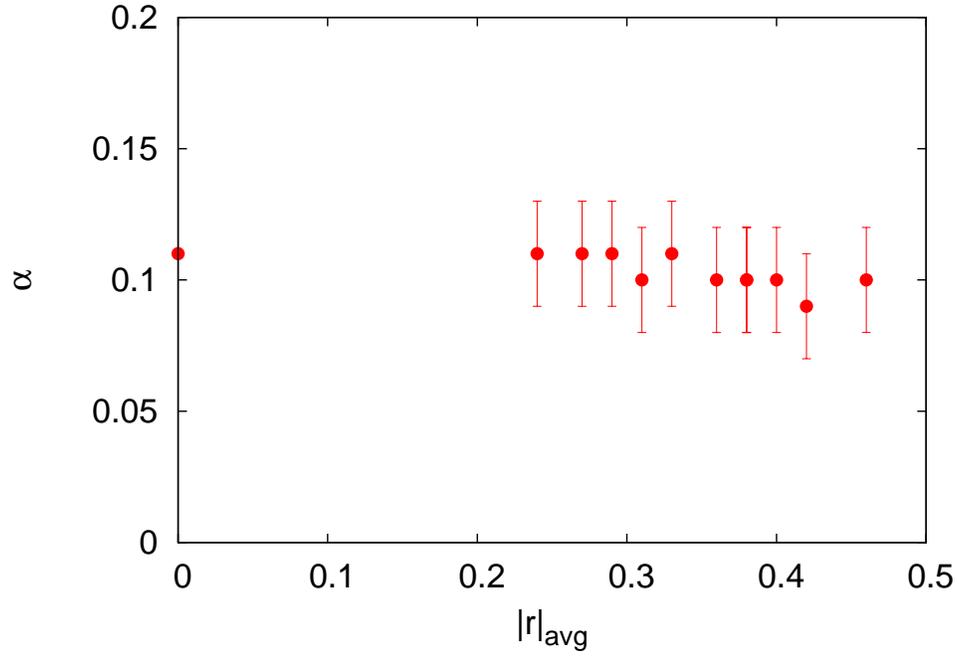}

\caption{(color online) Variation of the exponent $\alpha$ with 
$|r|_{avg}$ based on the same data and procedure as used to obtain 
figure (9). } ¬

\label{fig10}
\end{figure}

\end{document}